\documentclass[journal]{IEEEtran}

\usepackage{amsmath}
\usepackage[caption=false,font=footnotesize]{subfig}
\usepackage{cite}
\usepackage{graphicx}
\usepackage{epstopdf}
\usepackage{multirow}
\usepackage{pgfplots}
\usepackage{nicefrac}
\usepackage{comment}
\usepackage[normalem]{ulem}
\pgfplotsset{compat=newest}
\usepgfplotslibrary{polar}
\usetikzlibrary{arrows.meta}
\usepackage{tikz}
\usetikzlibrary{external}
\usepackage{siunitx}
\newlength\figureheight
\newlength\figurewidth
\newcommand{\figref}{Fig.~\ref}
\newcommand{\tabref}{Table~\ref}

\usepackage[absolute,overlay]{textpos}
\usepackage{isomath}
\usepackage{colortbl}
\newcolumntype{B}[1]{>{\centering\arraybackslash}b{#1}}

\usepackage{threeparttable}
\usetikzlibrary{plotmarks}
\usepackage{graphbox}

\begin{document}

\title{Pattern and Polarization Diversity Multi-Sector Annular Antenna for IoT Applications}%

\author{Abel~Zandamela,~\IEEEmembership{Graduate Student~Member,~IEEE,}
Nicola~Marchetti,~\IEEEmembership{Senior~Member,~IEEE,}
Max~J.~Ammann,~\IEEEmembership{Fellow,~IEEE,}
and~Adam~Narbudowicz,~\IEEEmembership{Senior Member,~IEEE}%
\thanks{This work was supported by Science Foundation Ireland under Grant 18/SIRG/5612.
A. Zandamela, N. Marchetti, and A. Narbudowicz are with CONNECT Centre, Trinity College Dublin, The University of Dublin, Dublin 2, Ireland (email: \{zandamea, nicola.marchetti, narbudoa\}@tcd.ie).}
\thanks{A. Narbudowicz is also with the Department of Telecommunications and Teleinformatics, Wroclaw University of Science and Technology, Wroclaw 50-370, Poland.}
\thanks{Max J. Ammann is with the Antenna and High Frequency Research
Centre, School of Electrical and Electronic Engineering, Technological
University Dublin, D07ADY7 Dublin, Ireland (e-mail:
max.ammann@tudublin.ie).}}%

\maketitle

\begin{abstract}
This work proposes a small pattern and polarization diversity multi-sector annular antenna with electrical size and profile of $\boldsymbol{ka=1.2}$ and $\boldsymbol{0.018\lambda}$, respectively. The antenna is planar and comprises annular sectors that are fed using different ports to enable digital beamforming techniques, with  efficiency and gain of up to 78\% and 4.62 dBi, respectively. The cavity mode analysis is used to describe the design concept and the antenna diversity. The proposed method can produce different polarization states (e.g. linearly and circularly polarized patterns), and pattern diversity characteristics covering the elevation plane. Owing to its small electrical size, low-profile and diversity properties, the solution shows good promise to enable advanced radio applications like wireless physical layer security in many emerging and size-constrained Internet of Things (IoT) devices.
\end{abstract}


\IEEEpeerreviewmaketitle

\section{Introduction}\label{sec:intro}

\IEEEPARstart{T}{HE} ability of an antenna to alter its radiated pattern and polarization characteristics is a highly desirable property in many wireless systems\cite{ Qin2010, Haupt2013, Zhou2014, Costantine2015, Li2017, Hu2022, doi:https://doi.org/10.1002/9781119813910.ch3}. In that regard, the rapid  advance of the Internet of Things (IoT) technology creates an increasing need for small pattern and polarization diversity antennas to support cutting-edge applications like on/off-body communications, medical body area networks, localization, physical layer security, and others \cite{Liu2013bodycentric, IoTMag22,  doi:https://doi.org/10.1002/9781119813910.ch9, Daly2010, YDing2018, Zandamela2023, Zhu2016}. 

Small pattern diversity antennas have been proposed, e.g., in \cite{Zhu2016,  Jin2018, Ouyang2018, Darvazehban2020, Wu2020, Wang2021, Jian2022, Zhao2022}. In \cite{Ouyang2018}, a unilateral turnstile-shaped patch antenna is proposed for azimuth plane beamsteering; the antenna has size $ka=1.32$, and profile $0.024\lambda$ (where $a$ is the radius of the smallest sphere that completely encloses the antenna at the center operating frequency $f=c/\lambda$, $k =
2\pi/\lambda$ is the free space wavenumber, and $\lambda$ is the wavelength). Magnetic and electric near-field resonant parasitic elements are used for pattern diversity in \cite{Wu2020} (size: $ka=0.98$, and profile: $0.0026\lambda$). A shared-aperture pattern diverse quasi-Yagi antenna with size $ka=3$ and a profile of $0.005\lambda$ is proposed in \cite{Wang2021}. A modified array structure (size: $ka=0.98$ and profile: $0.004\lambda$) is used to switch between unidirectional and omnidirectional patterns in \cite{Zhao2022}. Further studies in miniaturized antennas are presented, e.g., in \cite{Tang2018, Nnguyen2016, Chen2018, Zhang2021CM, Li2022, Li2020, Jia2022}. A quad-polarized Huygens dipole antenna with $ka=0.944$ and profile $0.044\lambda$ is discussed in \cite{Tang2018}. The theory of characteristics modes is used for quad-polarization in \cite{Zhang2021CM}, with size $ka=2.36$ and a profile of $0.008\lambda$. The work in \cite{Li2022} proposes a tri-polarized planar antenna with $ka=2$ and a profile of $0.04\lambda$. While the above-discussed designs present significant advances in efficiency, bandwidth, and single-diversity performance, some works either still have a relatively large electrical size or profile \cite{Zhu2016, Ouyang2018, Zhang2021CM, Li2020, Li2022, Chen2018}, or the pattern diversity is restricted to a few discrete states that cannot be activated simultaneously, e.g., in \cite{Jin2018, Ouyang2018, Darvazehban2020, Wu2020, Jian2022, Zhao2022, Tang2018, Nnguyen2016, Chen2018, Zhang2021CM, Li2022, Li2020}. This limits the implementation of many emerging and advanced wireless radio applications like localization and physical layer security techniques in small IoT devices. 

It should be highlighted that all the above works either allow for pattern or polarization diversity. Therefore, it is difficult to design a dual-diversity antenna with a small electrical size and low-profile; however, such antennas have the potential to enable many advanced wireless applications. In recent years, only a few works have proposed pattern and polarization diverse compact antennas, e.g., in \cite{Deng2017, Yang2019, Jia2022, Ramirez2021}. The work in \cite{Deng2017}, proposed a three-layer Yagi patch  antenna with $\pm\ang{45}$ dual-polarization and pattern diversity (size: $ka=5.2$ and profile $0.08\lambda$). In \cite{Yang2019}, a radiating patch antenna is combined with diagonal metal walls for multi-polarization and multi-directional beams (size: $ka=2$ and profile: $0.14\lambda$). In \cite{Jia2022}, a dual-polarized beamsteering active reflection metasurface antenna is proposed (size: $ka=4.68$, and profile $0.4\lambda$). Lastly, an aperture-stacked patch with cross-shaped parasitic strips is used for dual-polarization and pattern diversity in \cite{Ramirez2021} (size: $ka=5.3$ and profile: $0.23\lambda$). While the above designs can enable pattern and polarization diversity, their dimensions are still relatively large (i.e. $\geq2ka$) for size-constrained IoT applications. In addition, these structures are not planar, which can limit their integration in many emerging IoT systems. 
 
Microstrip patch antennas are a well-known solution for designing low-cost, low-profile, and planar antennas, which are some of the key requirements for IoT applications. For a microstrip patch antenna occupying an area that can be bounded by a region of radius $r$, one can design a sector microstrip patch by using only a portion of the patch area. The sector area is then given by $A_{sector}=\pi r^2({\alpha/2\pi})$, where $\alpha$ is the sector angle. Such structures can be advantageous over standard shapes like circular, annular rings, or rectangular shapes in terms of size and profile miniaturization, as well as simpler design structures. Sector microstrip antennas have been proposed, e.g., in \cite{Deshmukh2015, Liang2015, Liu2016, Lu2017, Wu2020_sector, Yu2020, Ghosh2021AWPL}. In \cite{Liang2015}, a driven circular sector and annular sector directors are used to design a quasi-Yagi array. In \cite{Liu2016}, a sector annular ring with a coupled sector patch is used for dual-band operation. Circular sector antennas are investigated for bandwidth enhancement and antenna miniaturization in \cite{Lu2017, Wu2020_sector, Yu2020, Ghosh2021AWPL}, with null-scanning in \cite{Wu2020_sector}, tilted circular polarization in \cite{Yu2020}, and low cross-polarization in \cite{Ghosh2021AWPL}. However, pattern and polarization diversity are not realized in the above designs, limiting their use for advanced radio applications in emerging small IoT devices.

In this work, we propose for the first time a small pattern and polarization diversity multi-sector annular microstrip patch antenna. The design comprises four-ports and operates near $\SI{4}{GHz}$, with electrical size $ka=1.2$, low-profile of $0.018\lambda$, and a $\SI{10}{dB}$ and $\SI{6}{dB}$ Impedance Bandwidth (IBW) of $\SI{11}{MHz}$ and $\SI{79}{MHz}$, respectively. The antenna is planar and exploits annular sectors, slits, and vias-based mutual coupling enhancement methods to realize a simpler design configuration with different polarization, e.g., linear, circular, and $\pm\ang{45}$ dual-polarization. In addition, good beamsteering characteristics covering the elevation plane are also demonstrated with an antenna gain of up to $\SI{4.62}{dBi}$. At first, the design principle of an annular sector antenna is presented. Then a multiport design is derived, and the mutual coupling enhancement techniques are discussed. Next, the diversity characteristics are described. Finally, experimental results are presented to validate the proposed design concept. 


\section{Working Principle and Antenna Configuration}
\label{sec:designPrinciple}%

\subsection{Design Principle}

For an annular sector microstrip antenna printed on an electrically thin substrate, i.e., $t\ll \lambda$ (where $t$ is the substrate thickness), the cavity model can be used to determine the generated electric fields at a point $(\rho,\phi)$ \cite{Richards1984}

\begin{equation}
\begin{aligned}
\boldsymbol{E} (\rho,\phi) = j\omega \mu \hat{z} \sum_{m,n} \frac{\psi_{mn} (\rho,\phi) \psi_{mn}(\rho',\phi')}{k^2 - k_{mv}^2} 
\end{aligned}
\label{eq:E_fields}
\end{equation}

\noindent where $k$ is the wavenumber, $(\rho',\phi')$ is the feed point in polar coordinates, $\mu$ is the permeability of the medium, and the eigenfunctions are computed as \cite{Lo1979}

\begin{figure}[!t]
\centering
\includegraphics[]{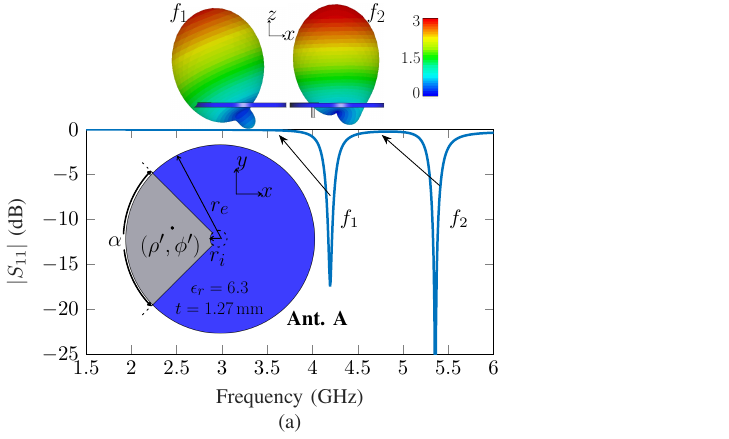}
\hspace{1.5cm}
\includegraphics[clip, trim=0cm 0cm 0cm 0cm, width=.42\textwidth]{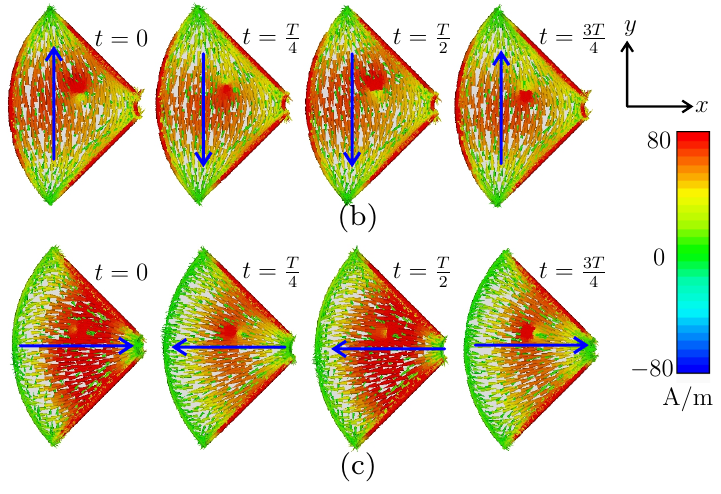}
\vspace{-.25cm}
\caption{Annular sector microstrip antenna: (a) S-parameters and 3D patterns of the first two modes of the proposed annular sector antenna [(Ant. A) shown as inset, with the color gray used for the metalization layer and the color blue represents the slots (no metalization, i.e. the substrate material)]; (b) surface current distribution of the annular sector at different time points for $f_{1}$ (the fundamental mode, i.e. mode 1); and (c) surface currents at $f_{2}$ (mode 2).}
\label{fig:patch_A0}
\end{figure}

\vspace{-.5cm}
\begin{equation} 
\psi_{mn} = \left[ J_v (k_{mv}\rho)N'_v(k_{mv}r_i) - J'_v(k_{mv}r_i)N_v(k_{mv}\rho) \right] \cos{v \phi}
\label{eq:eigen_funct}
\end{equation}

\noindent where $J_v$ and $N_v$ represent the cylindrical functions of the first and second kind (Bessel and Newman functions) of order $v$, respectively; $v=n\pi/\alpha$, $\alpha$ is the sector angle, $k_{mv}$ are the resonant wave numbers, which are solutions of 

\vspace{-.5cm}
\begin{equation}
J'_v (k_{mv}r_i)N'_v(k_{mv}r_e) = J'_v (k_{mv}r_e)N'_v(k_{mv}r_i) 
\label{eq:eigen_funct_solut}
\end{equation}

\noindent where $r_i$ and $r_e$ are the annular sectors' inner and outer radii, respectively. The corresponding resonance frequency for the specific mode can then be approximated by $f={ck_{mv}}/({{2\pi r_e \sqrt{\epsilon_{r}}}})$, where $c$ is the light velocity in vacuum, and $\epsilon_{r}$ is the substrate relative permittivity.

The inset in \figref{fig:patch_A0}a shows Ant. A, a simple annular sector antenna with $\alpha=\pi/2$, $n=1$, $r_i=\SI{1.5}{mm}$, and $r_e=\SI{14}{mm}$. The sector is printed on a TMM6 substrate ($\epsilon_r=6.3$, tan$ \delta=0.0023$, and thickness $t=\SI{1.27}{mm}$). The S-parameters are shown in \figref{fig:patch_A0}a, where the fundamental mode resonates at $f_{1}=\SI{4.2}{GHz}$, and the second mode at $f_{2}=\SI{5.35}{GHz}$. The surface current distribution at different time points is shown in \figref{fig:patch_A0}b (for mode 1) and \figref{fig:patch_A0}c (for mode 2), with the current direction highlighted for easier observation. It is noted that for mode 1 (at $f_1$), strong currents are seen at the center of the sector, and they are basically in the $y$ direction. At phase angle $t=\ang{0}$, the flowing direction is towards $+y$, while they flow towards $-y$ for the next two-quarters of the period and return to $+y$ direction for phase angle $t=\frac{3T}{4}$; this configuration then produces a linearly polarized broadside pattern with the main beam in the $xz-$plane [\figref{fig:patch_A0}a (top)]. For mode 2 (at $f_2$), the currents are along the $x$ direction. It is seen that the currents are stronger near the center and closer to the edges of the inner radius; at $t=\ang{0}$ strong currents are seen at the center, and the flow is towards the $+x$ direction; they then flow towards the $-x$ direction for the next two-quarters of the period reaching another state with strong currents at $t=\frac{T}{2}$, while returning to $+x$ for $t=\frac{3T}{4}$. This indicates that strong currents are seen at every one-quarter of a period with a change in direction of the currents, which produces a linearly polarized beam pattern but with the main beam pointing at broadside $\theta=\ang{0}$. In both cases: $|S_{11}|> \SI{16}{dB}$, and the total efficiency is $>85\%$. This performance is realized with total dimensions: $\SI{28}{mm} \times \SI{28}{mm} \times \SI{1.34}{mm}$ or $0.39\lambda \times 0.39\lambda \times 0.018\lambda$, where $\lambda$ is the wavelength at $f_{1}=\SI{4.2}{GHz}$. 


\subsection{Multiport Design} 
 
In the next design step, three additional annular sectors are integrated into the antenna volume. The total electric field of the antenna at $r>8r_e^2/\lambda$ can then be approximated as the superposition of the $L=4$ annular sectors using

\vspace{-.25cm}
\begin{equation}
\boldsymbol{E}_{tot} (r,\theta,\phi)= \sum_{l=1}^{L}c_l \boldsymbol {E}_{l}
\label{eq:total_pat}
\end{equation}

\noindent where $c_l=|A_l|e^{j\Delta\beta_l}$ is the excitation coefficient of the $l$th annular sector; $|A_l|$, $\Delta\beta_l$ are the amplitude and phase excitation, respectively; $\boldsymbol{E}_l$ is $l$th sector electric field obtained from \eqref{eq:E_fields}.

\begin{figure}[!t]
\centering
\subfloat[]{
\hspace{-.5cm}
\includegraphics[]{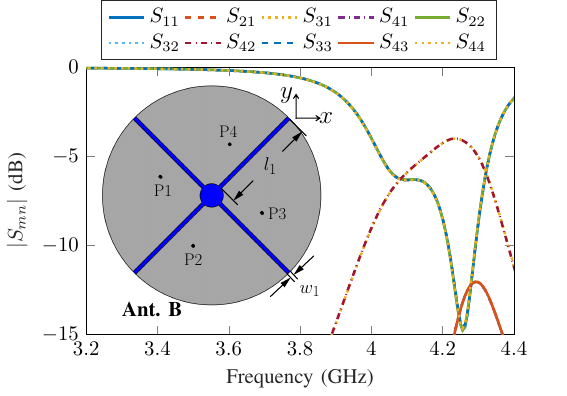}
\label{fig:patch_A1}}
\hfill
\vspace{-.15cm}
\subfloat[]{
\hspace{-.5cm}
\includegraphics[]{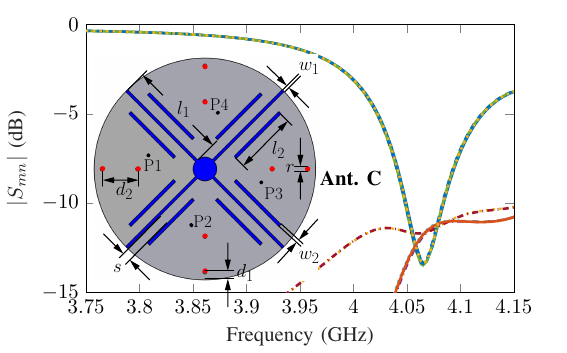}
\label{fig:patch_A2}}
\caption{S-parameters of the proposed antennas: (a) Ant. B (shown as inset), obtained using the basic structure of Ant. A operating at $f_{1}$ mode; and (b) Ant. C operating at $f_{1}$ mode with electrical size $ka=1.2$.}
\end{figure}

To allow separation between the annular sectors, four slits of width $w_1=\SI{0.5}{mm}$, and length $l_1=\SI{12.5}{mm}$ are used (see \figref{fig:patch_A1}). The coordinates of the feed point are P1$(x,y)=$~P1$(\SI{-6.5}{mm},\SI{2.3}{mm})$, and the feed of the remaining ports are obtained by rotating P1 by $\ang{90}$ with respect to the center of the substrate. The mutual coupling $|S_{mn}|$ between different sectors is obtained from the finite element method 3D full-wave solver. For the multiport Ant. B (inset of \figref{fig:patch_A1}), which uses the basic structure of Ant. A, the $|S_{mn}|<\SI{4}{dB}$ and are shown in \figref{fig:patch_A1}. Such mutual coupling values are considered insufficient for many applications, as they can significantly reduce the overall system performance, e.g., antenna gain and beamsteering characteristics. 

To improve $|S_{mn}|$, a reactive loading in the form of shorting pins is introduced, which also allows for improved frequency tuning. From Ant. B, a single pin of radius $r$ is placed at $d_1=\SI{1}{mm}$ from the edge of the outer diameter (see \figref{fig:patch_A2}); because of the proximity to the edge, this value should be adjusted for larger pin radius. Next, different $r$ values $0.125 - \SI{0.5}{mm}$ are tested and $|S_{mn}|$ stays below $\SI{5.2}{dB}$ in all cases. The value $d_1$ is also tested from $1 -\SI{3}{mm}$ (with $r=\SI{0.25}{mm}$), and the $|S_{mn}|$ also stays below $\SI{5.2}{dB}$. In the next step, a second pin also of radius $r=\SI{0.25}{mm}$ is introduced at distance $d_2$ (see \figref{fig:patch_A2}); this value is tested from $1-\SI{7}{mm}$, and $d_2=\SI{4.5}{mm}$ with $|S_{mn}|<\SI{6.2}{dB}$ is chosen. Further mutual coupling enhancement is realized by tuning the feed position, tuning the gap between each sector $(w_1)$, and adding slits of length $(l_2)$ and width $(w_2)$. The slits are placed at distance $s$ from the sector side; also note that the surface currents' path will increase as they flow around the slit, providing miniaturization. For further size reduction, a second slit with a similar configuration is introduced at the opposite side of the sector. Note that techniques like high-permittivity loading can also be used with the proposed antenna. However, for lower $ka$ values, increased antenna losses are observed, with decreased total efficiency and deterioration of the coupling characteristics.

\begin{figure*}[!t]
\begin{center}
\subfloat[]{
\includegraphics[]{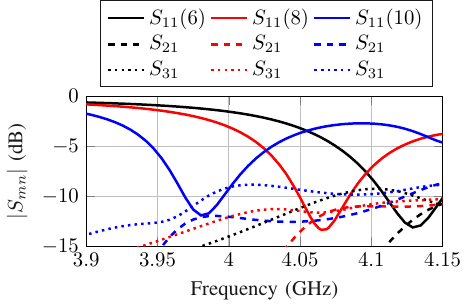}
\label{fig:slit_Length}}
\subfloat[]{
\includegraphics[]{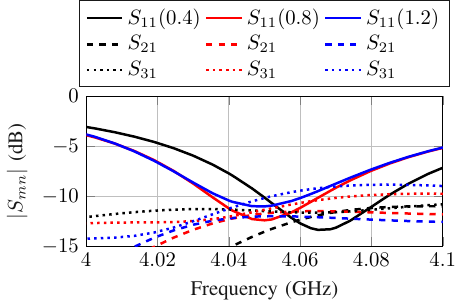}
\label{fig:slit_Width}}
\hfill
\subfloat[]{
\includegraphics[]{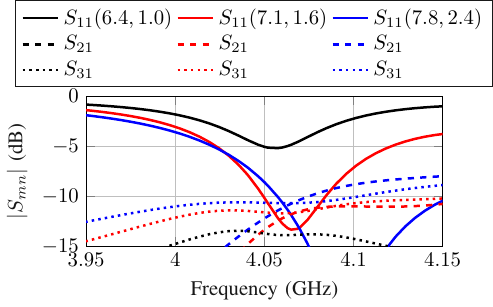}
\label{fig:FeedPos}}
\subfloat[]{
\includegraphics[]{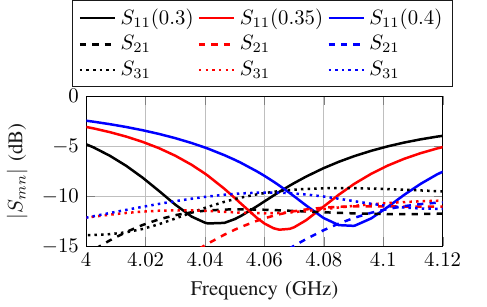}
\label{fig:w1_Parametric}}
\caption{Full-wave simulated parametric results for (all values in mm): (a) slit length ($l_2$); (b) slit width ($w_2$); (c) feed position; and (d) sectors gap $(w_1)$.}
\label{fig:param_studies}
\end{center}
\end{figure*}

\figref{fig:param_studies} shows parametric studies of Ant. C. For brevity, only $|S_{11}|$ and the coupling between the opposite sector $|S_{31}|$ and one adjacent sector $|S_{21}|$ are shown. \figref{fig:slit_Length} shows the $|S_{mn}|$ for different $l_2$ values for $s=\SI{1.3}{mm}$. It is seen that $|S_{mn}|>\SI{11}{dB}$ is realized for $l_2=\SI{8}{mm}$, and the center frequency increases for lower $l_2$ values. The parametric results for $w_2$ are shown in \figref{fig:slit_Width}; it is observed that $w_2=\SI{0.4}{mm}$ achieves better isolation characteristics. \figref{fig:FeedPos} shows the results for different feed positions along the slit length, and the optimal feed location is $(\SI{7.1}{mm}, \SI{1.6}{mm})$. The gap between each sector is also tested for different values and the results are shown in \figref{fig:w1_Parametric}. It can be seen that the center frequency increases for larger $w_1$ values, and $|S_{mn}|>\SI{11}{dB}$ is realized for $w_1=\SI{0.35}{mm}$. 
The $|S_{mn}|$ with optimized parameters are shown in \figref{fig:patch_A2}, the frequency is lowered from $f_1=\SI{4.25}{GHz}$ (Ant.B) to $f_1=\SI{4.065}{GHz}$ (Ant. C), where the total efficiency is $>60\%$ and $|S_{mn}|>\SI{11.7}{dB}$. The $\SI{10}{dB}$ and $\SI{6}{dB}$ impedance bandwidths (IBW) are respectively $\SI{36.5}{MHz}$ and $\SI{82}{MHz}$, and the total efficiency is $>40\%$ across the entire $\SI{6}{dB}$ impedance bandwidth region. The antenna dimensions are $0.38\lambda \times 0.38\lambda \times 0.018\lambda$, and $ka=1.2$.


\section{Pattern and Polarization Diversity}
\label{sec:antenna_reconfig}%

\begin{figure}[!t]
\centering
\includegraphics[clip, trim=0cm 0cm 0cm 0cm, width=.46\textwidth]{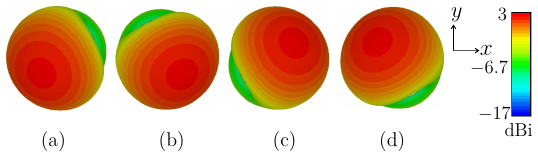}
\caption{3D realized gain for Ant. C: (a) P1; (b) P2; (c) P3; and (d) P4.}
\label{fig:PortsPattern}
\end{figure}

\begin{figure}[!t]
\vspace{-.5cm}
\captionsetup[subfloat]{labelformat=empty}
\centering
\subfloat[]{
\hspace{-1.1cm}
\includegraphics[]{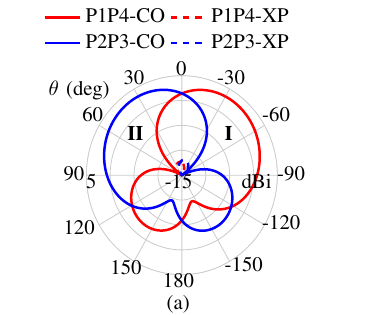}
\label{fig:P1P4_P2P3}}
\subfloat[]{
\hspace{-2cm}
\includegraphics[]{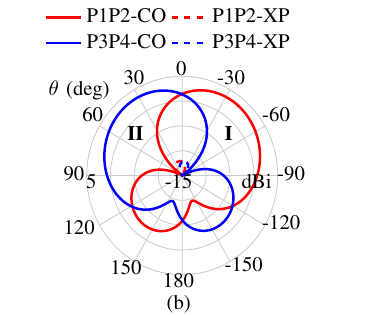}
\label{fig:P1P2_P3P4}}
\vspace{-.5cm}
\caption{Pattern diversity of the proposed antenna: (a) linearly polarized broadside patterns covering $xz$-plane ($\phi=\ang{0}$); and (b) $yz$-plane ($\phi=\ang{90}$).}
\end{figure}

\figref{fig:PortsPattern} shows the radiation pattern for each sector of the proposed antenna, and the peak realized gain is around $\SI{2.64}{dBi}$ for each case. Using \eqref{eq:total_pat}, different states are created to obtain pattern and polarization diversity. When P1 and P4 are simultaneously excited with no additional phase-shifts, i.e. P1 ($|A_1|e^{j\Delta\beta_1}=|A_1|$) and P4 ($|A_4|e^{j\Delta\beta_4}=|A_4|$), sectors 1 and 4 generate a radiation pattern with the main beam in the $xz-$plane Quadrant I (see \figref{fig:P1P4_P2P3}). When P2 and P3 are excited simultaneously, they produce a pattern with the main beam in the $xz-$plane Quadrant II, or at $\ang{70}$ with respect to the main beam of P1 and P4 (i.e., $\theta_{P2P3}= \theta_{P1P4} + \ang{70}$). To cover the $yz$-plane ($\phi=\ang{90}$), P1 ($|A_1|e^{j\Delta\beta_1}=|A_1|$) and P2 ($|A_2|e^{j\Delta\beta_2}=|A_2|$) are excited simultaneously to create a pattern with the main beam in the $yz-$plane Quadrant I (see \figref{fig:P1P2_P3P4}); and a second pattern in Quadrant II, can be generated by simultaneously exciting P3 and P4.

\begin{table*}[!t]
\caption{Input Amplitudes and Phase Configurations for Pattern and Polarization Diversity of The Proposed Antenna}
\vspace{-0.4cm}
\hspace{-0.3cm}
\centering
\begin{threeparttable}
\begin{tabular}{l||c|c|c|c|c|c|c|c|c|c}
\hline
\label{tab:pattern_reconf}

Beam direction & Polariz. & Gain $^{*}$(dBi) & P1 ($|A_{1}|$) & P1 ($\Delta\beta_1$) & P2 ($|A_{2}|$) & P2 ($\Delta\beta_2$) & P3 ($|A_{3}|$) & P3 ($\Delta\beta_3$) & P4 ($|A_{4}|$) & P4 ($\Delta\beta_4$) \\ \hline\hline

$\theta=\ang{-35}, \phi=\ang{0}$  & LP  & $3.44$ &  $1$   &  $\ang{0}$  &  -  &   - &  -  & - &  $1$  & $\ang{0}$     \\ \hline

$\theta=\ang{35}, \phi=\ang{0}$  & LP  & $3.44$ &  -   &  -  &  1  &   $\ang{0}$ &  $1$  & $\ang{0}$ &  -  & -    \\ \hline

$\theta=\ang{-35}, \phi=\ang{90}$  & LP  & $3.47$ &  $1$   &  $\ang{0}$  &  $1$ &   $\ang{0}$ &  -  & - &  -  & -     \\ \hline

$\theta=\ang{35}, \phi=\ang{90}$  & LP  & $3.47$ &  -  &  -  &  - &   - &  $1$  & $\ang{0}$ &  $1$ & $\ang{0}$     \\ \hline

Omnidirectional  & HP  & $0.23$ &  $1$   &  $\ang{0}$  &  $1$ &   $\ang{0}$ &  $1$  & $\ang{0}$ &  $1$  & $\ang{0}$    \\ \hline

$\theta=\ang{0},\phi=\ang{0}$  & LP  & $4.59$ &  $1$   &  $\ang{0}$  &  $1$ &   $\ang{0}$ &  $1$  & $\ang{180}$ &  $1$  & $\ang{180}$    \\ \hline

$\theta=\ang{0},\phi=\ang{-45}$  & DP  & $4.59$ &  $1$   &  $\ang{0}$  &  - &  - &  $1$  & $\ang{180}$ &  - & -    \\ \hline

$\theta=\ang{0},\phi=\ang{45}$  & DP  & $4.59$ &  -   & -  &  $1$ &  $\ang{0}$ &  -  & - &  $1$ & $\ang{180}$    \\ \hline

$\theta=\ang{0},\phi=\ang{0}$  & RHCP  & $4.56$ &  $1$   & $\ang{0}$  &  $1$ &  $\ang{90}$ &  $1$  &  $\ang{180}$  &  $1$ & $\ang{270}$    \\ \hline

$\theta=\ang{0},\phi=\ang{0}$  & LHCP  & $4.56$ &  $1$   & $\ang{270}$  &  $1$ &  $\ang{180}$ &  $1$  &  $\ang{90}$  &  $1$ & $\ang{0}$    \\ \hline
\end{tabular}
\begin{tablenotes}[para,flushleft]
\centering $^{*}$ Realized gain, $|A_l|$ (amplitude), $\Delta\beta_l$ (phase shift), LP (linear polarization), HP (horizontally polarized), DP ($\pm \ang{45}$ dual-polarization). 
\end{tablenotes}
\end{threeparttable}
\end{table*}

\begin{figure}[!t]
\captionsetup[subfloat]{labelformat=empty}
\centering
\subfloat[]{
\hspace{-1cm}
\includegraphics[]{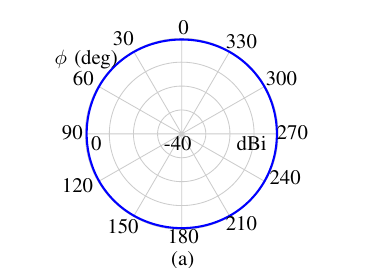}}
\subfloat[]{
\hspace{-2cm}
\includegraphics[]{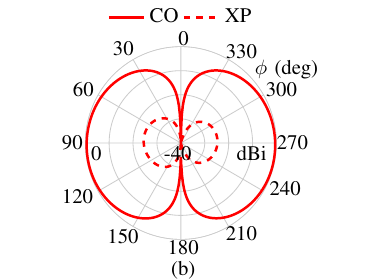}}
\vspace{-.5cm}
\caption{Proposed antenna horizontally polarized omnidirectional pattern: (a) $\theta=\ang{90}$ plane; and (b) $\phi=\ang{0}$ plane.}
\label{fig:monop_reconf}
\end{figure}

\begin{figure}[!t]
\vspace{-.5cm}
\captionsetup[subfloat]{labelformat=empty}
\centering
\subfloat[]{
\hspace{-1.25cm}
\includegraphics[]{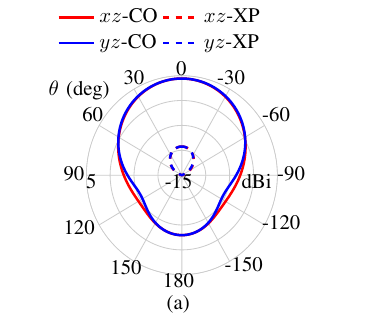}
\label{fig:broad_lp}}
\subfloat[]{
\hspace{-2.2cm}
\includegraphics[]{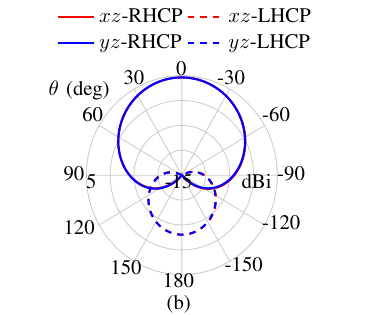}
\vspace{-1cm}
\label{fig:rhcp_reconf}}
\vspace{-.5cm}
\caption{Polarization diversity of the proposed antenna: (a) linearly polarized patterns for $xz$- and $yz$-planes; and (b) RHCP pattern for the same cuts.}
\end{figure}

\begin{figure}[!t]
\centering
\subfloat[]{
\hspace{0cm}
\includegraphics[]{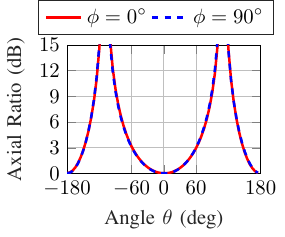}}
\subfloat[]{
\hspace{-.5cm}
\includegraphics[]{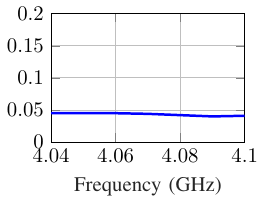}}
\caption{Axial ratio of the proposed antenna: (a) axial ratio for $xz$- and $yz$-planes at $f_1=\SI{4.065}{GHz}$; and (b) axial ratio over frequency.}
\label{fig:ARsimul}
\end{figure}

\figref{fig:monop_reconf} shows a horizontally polarized omnidirectional pattern obtained by simultaneous excitation of all the antenna ports without additional phase shifts in each excited port. \figref{fig:broad_lp} shows a linearly polarized pattern with the main beam pointing at $\theta=\ang{0}$. The pattern is generated by exciting pairs (P1, P3) and (P2, P4) with a $\ang{180}$ phase difference between the ports; this is ($|A_1|e^{j\Delta\beta_1}=|A_1|$,~$|A_3|e^{j\Delta\beta_3}=-|A_3|$), and ($|A_2|e^{j\Delta\beta_1}=|A_2|$,~$|A_4|e^{j\Delta\beta_4}=-|A_4|$). It is also interesting to note that the excitation of only one pair will produce a $\ang{\pm 45}$ polarization. In this case exciting only P1 and P3 (with $\ang{180}$ phase difference) will produce a broadside pattern with $-\ang{45}$ polarization and the pair P2 and P4 (with $\ang{180}$ phase difference) will generate a $+\ang{45}$ polarization. 

The results in \figref{fig:rhcp_reconf} also show the polarization diversity of the proposed antenna. To realize circular polarization, a sequential feed with $\ang{90}$ phase difference between each adjacent port is used, i.e. P1 ($|A_1|, \Delta\beta_1=\ang{0}$), P2 ($|A_2|, \Delta\beta_2=\ang{90}$), P3 ($|A_3|, \Delta\beta_3=\ang{180}$), and P4 ($|A_4|, \Delta\beta_4=\ang{270}$). The configuration produces a broadside Right-Hand CP (RHCP) pattern (see \figref{fig:ARsimul}a). The axial ratio of the obtained patterns is $<\SI{1}{dB}$ within the entire $\SI{6}{dB}$ IBW (see \figref{fig:ARsimul}b). Note that a Left-Hand CP (LHCP) pattern can be obtained by reversing the above-presented phase configuration. Lastly, Table I outlines each port excitation value to generate the pattern and polarization diversity characteristics of the proposed design.


\section{Experimental Results} 

\begin{figure}[!t]
\centering
\subfloat[]{
\hspace{-0.3cm}
\includegraphics[clip, trim=0cm 0cm 0cm 0cm, align=c, width=.15\textwidth]{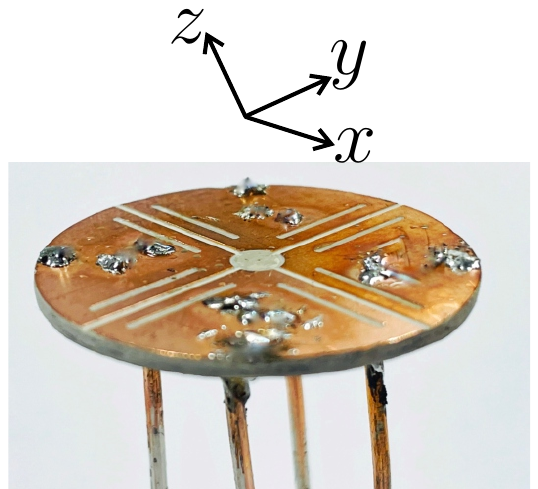}}
\subfloat[]{
\includegraphics[clip, trim=0cm 0cm 0cm 0cm, align=c, width=.162\textwidth]{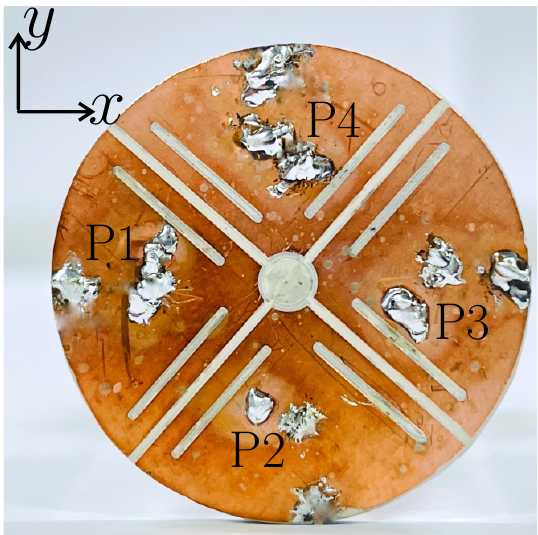}}
\subfloat[]{
\includegraphics[clip, trim=3cm 0cm 1.75cm 1cm, align=c, width=.163\textwidth]{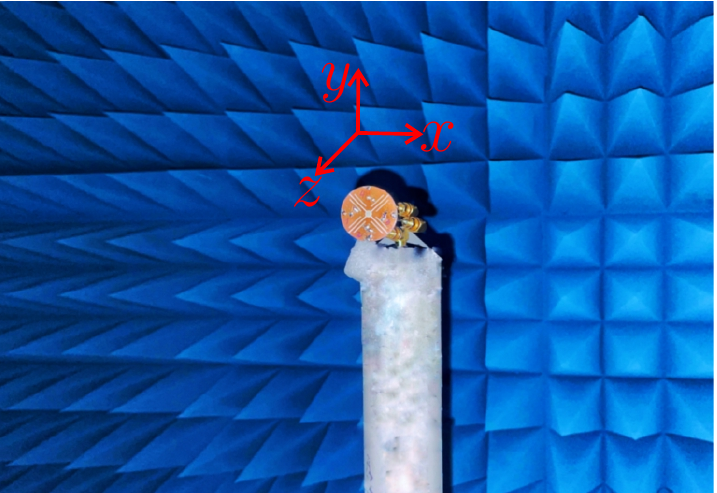}}
\caption{Manufactured prototype: (a) perspective view; (b) top view; and (c) anechoic chamber measurement setup.}
\label{fig:ant_prototype}
\end{figure}

\begin{figure}[!t]
\centering
\hspace{.5cm}
\includegraphics[]{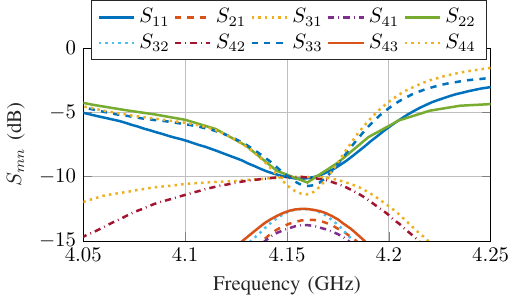}
\caption{Measured S-parameters of the manufactured antenna shown in \figref{fig:ant_prototype}.}
\label{fig:meas_spara}
\end{figure}

To validate the proposed concept, the antenna was manufactured using LPKF ProtoMat M60 and is shown in \figref{fig:ant_prototype}. The S-parameters measurements were conducted using a four-port VNA (R\&S ZVA40). It can be seen that the center frequency shifted upwards from $\SI{4.065}{GHz}$ (simulated case, \figref{fig:patch_A2}) to $\SI{4.16}{GHz}$ (see \figref{fig:meas_spara}). The $\SI{10}{dB}$ IBW changes from $\SI{36.5}{MHz}$ in simulations to $\SI{11}{MHz}$ in measurements; while the $\SI{6}{dB}$ IBW changes from $\SI{82}{MHz}$ in simulations, to $\SI{79}{MHz}$ in measurements. These discrepancies are most likely due to manufacturing and permittivity tolerances. 

\begin{figure}[!t]
\vspace{-.75cm}
\captionsetup[subfloat]{labelformat=empty}
\centering 
\subfloat[]{
\hspace{-1.7cm}
\includegraphics[]{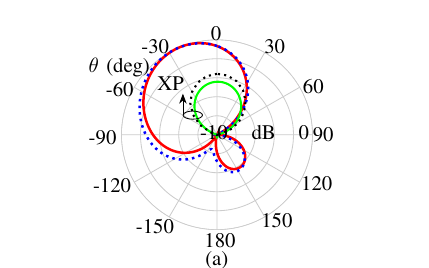}}
\subfloat[]{
\hspace{-3cm}
\includegraphics[]{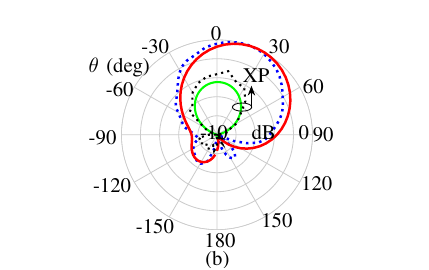}}
\hfill
\vspace{-.9cm}
\subfloat[]{
\hspace{-1.7cm}
\includegraphics[]{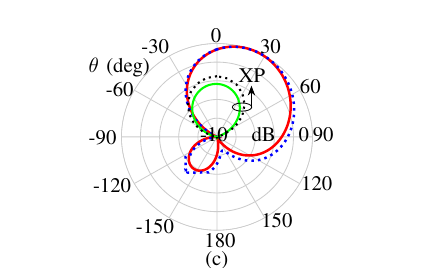}}
\subfloat[]{
\hspace{-3cm}
\includegraphics[]{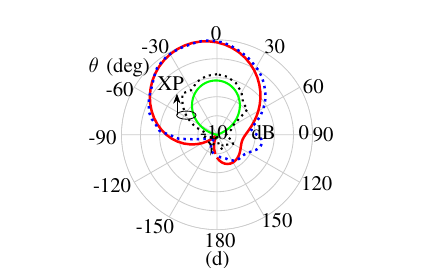}}
\vspace{-.5cm}
\caption{Measured (dotted lines) and simulated (solid-lines) normalized patterns of the proposed antenna in the $xz$-plane ($\phi=\ang{0}$) for each antenna port: (a) P1; (b) P2; (c) P3; and (d) P4.}
\label{fig:meas_ports}
\end{figure}

Anechoic chamber measurements were conducted and the measurement setup is shown in \figref{fig:ant_prototype}c. The measured patterns ($xz$-plane, $\phi=\ang{0}$) are shown in \figref{fig:meas_ports} at $f_1=\SI{4.16}{GHz}$ for each respective port. Overall, it is seen that the measured patterns have comparable properties with the respective simulated cases, with their main beams pointing towards: $\theta=\ang{-23}$ (in simulations) and $\theta=\ang{-25}$ (in measurements) for port 1 as shown in \figref{fig:meas_ports}a; $\theta=\ang{21}$(in simulations) and $\theta=\ang{19}$ in the measured case, for port 2 (see \figref{fig:meas_ports}b); for port 3, $\theta=\ang{23}$ for both the simulated and measured cases (\figref{fig:meas_ports}c); and finally for port 4, $\theta=\ang{-21}$ for both the simulated and measured cases (\figref{fig:meas_ports}d). The small beamwidth discrepancies seen in the measured patterns, the main beam direction for ports 1 and 2, are most likely due to antenna holder reflections during the measurements (see \figref{fig:ant_prototype}c). 

\begin{figure}[!t]
\centering
\hspace{.5cm}
\includegraphics[]{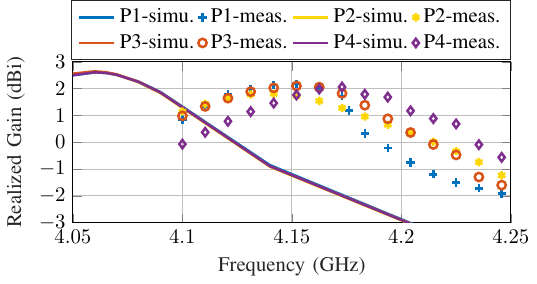}
\caption{Realized gain over frequency of each port of the proposed antenna for simulated results (solid lines) and measured results (shown using markers).}
\label{fig:RealGain}
\end{figure}

\begin{figure}[!t]
\vspace{-.75cm}
\captionsetup[subfloat]{labelformat=empty}
\centering 
\subfloat[]{
\hspace{-1.7cm}
\includegraphics[]{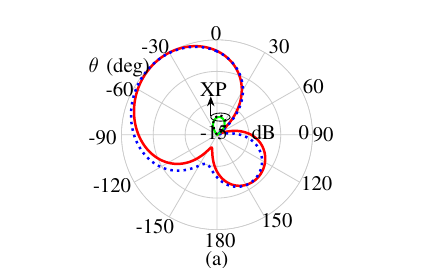}}
\subfloat[]{
\hspace{-3cm}
\includegraphics[]{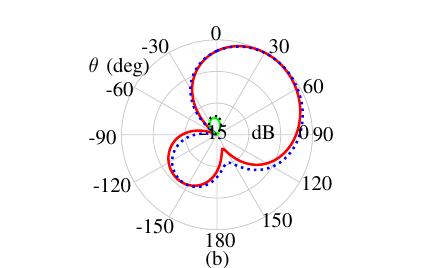}}
\vspace{-.5cm}
\caption{Measured (dotted lines) and simulated (solid-lines) normalized patterns for $\phi=\ang{0}$ plane: (a) $\theta=\ang{-35}$ obtained from P1 and P4; and (b) $\theta=\ang{35}$ generated from P2 and P3.}
\label{fig:meas_p1p4_p2p3}
\end{figure}

\figref{fig:RealGain} shows the comparisons between the measured and simulated realized gain for each port. At the center frequency of each case (i.e., $\SI{4.065}{GHz}$ in simulations and $\SI{4.16}{GHz}$ in measurements), the peak realized gain is $\SI{2.64}{dBi}$ in simulations, and it slightly decreases to $\SI{2.21}{dBi}$ in measurements. Such discrepancy may be attributed to manufacturing and substrate tolerances and reflections due to the antenna holder during measurements. \figref{fig:meas_p1p4_p2p3} shows pattern diversity in the $xz$-plane, \figref{fig:meas_p1p4_p2p3}a shows the radiation pattern for $\theta=\ang{-35}$ direction, and the $\theta=\ang{35}$ radiation pattern is shown in \figref{fig:meas_p1p4_p2p3}b. Overall, good agreement is obtained between the simulated and measured cases, and the beamwidth differences may be explained by the beamwidth discrepancies highlighted in \figref{fig:meas_ports}. 

\begin{figure} [!t]
\vspace{-.75cm}
\captionsetup[subfloat]{labelformat=empty}
\centering 
\subfloat[]{
\hspace{-1.7cm}
\includegraphics[]{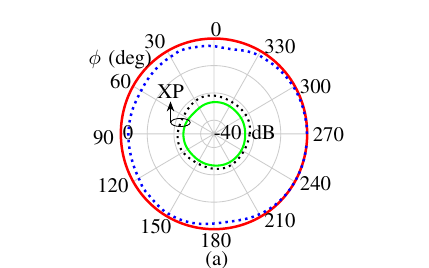}}
\subfloat[]{
\hspace{-3cm}
\includegraphics[]{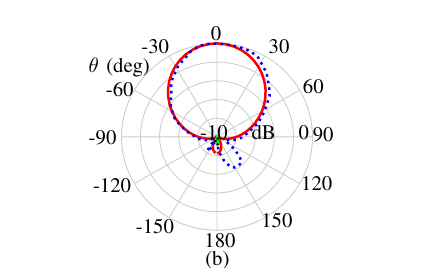}}
\vspace{-.5cm}
\caption{Measured (dotted lines) and simulated (solid-lines) normalized patterns for (a) linear horizontally polarized omnidirectional pattern ($\theta=\ang{90}$ cut-plane), and (b) linearly polarized broadside pattern for $xz$-plane.}
\label{fig:meas_polar_reconf}
\end{figure}

\figref{fig:meas_polar_reconf}a shows a horizontally polarized omnidirectional pattern. In this case, too, a good agreement is obtained between the simulated and measured cases, with the small shouldering and deeps most likely resulting from the beamwidth discrepancies seen in \figref{fig:meas_ports}. A linearly polarized broadside pattern is shown in \figref{fig:meas_polar_reconf}b. The results also demonstrate good agreement with their counterpart simulated values, validating the efficacy of the proposed design concept.

\begin{table*}[!t]
\caption{Comparison with Previously Published Works}
\label{tab:survey}
\centering
\begin{threeparttable}
\begin{tabular}{l||c|c|c|c|c|c|c|c|c}
\hline
Ref.             & $f_0$ (GHz)       & Size ($ka$) & Profile ($\lambda$) & Function & FBW($\%$) & Gain (dBi) & Effi. & Switches & Geometry \\ \hline\hline

\cite{Jin2018}   & $2.45$  & $2.72$ &  $0.0067$     & 4-D (Az)                  & $33.6$       & $4.11$        & $60\%$ & Yes  & Planar         \\ \hline

\cite{Ouyang2018}    & $2.44$      & $1.32$ & $0.024$      &  4-D (Az), (Ev) &   $15$ &  3.42        &  $83\%$ & Yes    & Planar        \\ \hline

\cite{Darvazehban2020} & $1.1$ & $1.74$ & $0.005$ & 3-D & $55$ & $5$ & - & Yes & Planar\\ \hline

\cite{Wu2020} & $1.51$ & $0.98$ & $0.0026$ & 1-D, 1-B & $1.32$ & 5.4 & $85\%$ & Yes & Planar\\ \hline

\cite{Wang2021} & $3.3$ & $3$ & $0.005$ & 1-O, 1-B, 2-T & $14.1$ & $7.38$ & $83.4\%$ & No & Planar\\ \hline

\cite{Zhao2022} & $2.45$ & $0.98$ & $0.004$ & 2-D, 1-O & $10.5$ & $6$ & $88.6\%$ & Yes & Planar\\ \hline

\cite{Nnguyen2016} & $0.402$ & $0.13$ & $0.0008$ & 3-D (Az) & $2.3$ & $-20$ & $0.4\%$ & Yes & Planar\\ \hline

\cite{Chen2018} & $2.45$ & $2.28$ & $0.28$ & 5-D (Ev) & $9.2$ &  $6.7$ & $90\%$ & Yes & Non-Planar \\ \hline

\cite{Tang2017} & $1.56$ & $0.92$ & $0.05$ &  3-D (Az) & $2.73$ & $3.59$ & $85\%$ & Yes & Non-Planar \\ \hline

\cite{Tang2018} & $1.5$ & $0.944$ & $0.045$ & LP,CP & $0.86$ & $2.97$ & $67\%$ & Yes & Planar\\ \hline

\cite{Zhang2021CM} & $2.45$ & $2.36$ & $0.008$ & LP,CP & $2.08$ & $4.22$ & - & Yes & Non-Planar\\ \hline

\cite{Li2022} & $2.35$ & $2$ & $0.004$ & LP,CP & $27.6$  & 6.39 & - & Yes & Planar\\ \hline

\cite{Li2020} & $4.8$ & $2.48$ & $0.037$ & LP,CP & $7.1$ & $6.63$ & - & Yes & Planar\\ \hline

\cite{Deng2017} & $3.5$ & $5.2$ & $0.08\lambda$ & 4-D (Ev), $\pm\ang{45}$DP & $5.6$ & $7.3$ & $70\%$ & Yes & Non-Planar\\ \hline

\cite{Ramirez2021} & $3.75$ & $5.3$ & $0.23\lambda$ & 3-D (Ev),  DP & $24$ & $7$ & $80\%$ & Yes & Non-Planar\\ \hline

\cite{Jia2022} & $3.45$ & $4.68$ & $0.4\lambda$ & 6-D, DP & $8.7$ & $>5$ & $88\%$ & Yes & Non-Planar\\ \hline

\cite{Yang2019} & $3.27$ & $2$ & $0.14$ & 6-D (Ev), LP, CP & $10.7$ & $6$ & $87\%$ & Yes & Non-Planar\\ \hline

\cellcolor{cyan}
{Prop.}   &  \cellcolor{cyan} ${4.16}$   &  \cellcolor{cyan} ${1.2}$ &  \cellcolor{cyan} $ {0.018}$      & \cellcolor{cyan}  {1-O, ${>8}$-D (Ev),LP,CP,${\pm 45^{\circ}}${DP}}  & \cellcolor{cyan}  $1.9^*$  & \cellcolor{cyan}  ${4.62}$ & \cellcolor{cyan}  ${78\%}$ & \cellcolor{cyan}  {No}      &  \cellcolor{cyan} {Planar}     \\ \hline \hline
\end{tabular}
\begin{tablenotes}[para,flushleft]
\centering $k=2\pi/\lambda$, $a$ is the radius of the smallest sphere that completely encloses the respective antenna, $^* \SI{6}{dB}$ FBW (fractional bandwidth), Az (Azimuth plane), Ev (elevation plane), B (Bidirectional), D (Directional), T (Tilted) patterns, 4-D (corresponds to four directional patterns), and symbol (-) for not given. 
\end{tablenotes}
\end{threeparttable}
\end{table*}

\begin{figure}[!t]
\centering
\hspace{0.525cm}
\includegraphics[]{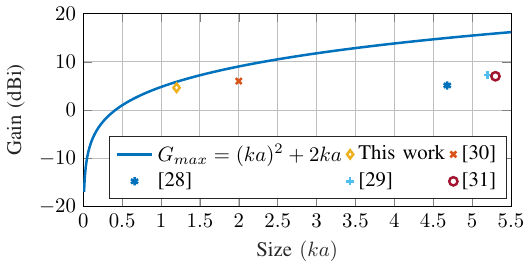}
\caption{Peak gain comparisons for the proposed antenna and previously published works against the Harrington maximum gain (computed with respect to the size of the sphere enclosing the antennas).}
\label{fig:TheorGain}
\end{figure}

To highlight the novelty and main advantages of the proposed antenna, a comparison with previously published works is outlined in \tabref{tab:survey}. The works in \cite{Jin2018, Ouyang2018, Darvazehban2020, Wu2020, Wang2021, Tang2017, Nnguyen2016, Chen2018, Tang2018, Zhang2021CM, Li2022, Li2020}, either allow for different beam patterns or polarization diversity. Even though the works in \cite{Deng2017, Ramirez2021, Jia2022, Yang2019} realize different beam patterns and polarization states, it can be seen that our proposed solution is planar and is capable of pattern and polarization diversity with a structure of only $ka=1.2$ and $0.018\lambda$ profile. To further complement the above comparisons, \figref{fig:TheorGain} shows the Harrington maximum theoretical gain ($G_{max}$) computed with respect to the size of the smallest sphere that fully encloses each proposed antenna \cite{Harrington1958}. It is observed that the multi-sector design realizes $\SI{4.62}{dBi}$ gain; therefore, it quickly approaches its maximum theoretical gain of $\SI{5.84}{dBi}$ as compared to other previously published pattern and polarization diversity antennas,  making it a good candidate to enable advanced radio applications in emerging small Internet of Things devices.

\section{Conclusion}\label{sec:conc}%

A very low-profile ($0.018\lambda$), compact size ($ka=1.2$), pattern, and polarization diversity annular sector-based antenna, with a gain close to its Harrington maximum gain, was presented. The diversity characteristics are realized by analysis of the modes excited in an annular sector antenna. By using a simple mutual coupling enhancement technique based on slits and vias loading, four concentrically rotated annular sectors are excited using four different ports. The proposed method realizes pattern diversity covering the elevation plane and generates a horizontally polarized omnidirectional pattern, a broadside pattern with linear, circular, and a $\pm\ang{45}$ dual-polarization. A prototype was manufactured and tested, and a good agreement was demonstrated between the measured and simulated results, validating
the design concept. The diversity characteristics were achieved using a planar structure designed on a single printed layer without requiring externally controlled switches. The structure is simple and compact and supports pattern and polarization diversity with simultaneous excitation of multiple beams within a structure requiring only $ka=1.2$. Therefore, it offers performance and size to allow advanced wireless radio applications in emerging size-constrained Internet of Things devices.

\bibliographystyle{IEEEtran}
\bibliography{Ref}

\ifCLASSOPTIONcaptionsoff
  \newpage
\fi

\end{document}